\documentclass[twocolumn,superscriptaddress,
amsmath,amssymb,pra,longbibliography,reprint]{revtex4-2}

\usepackage{listings}
\usepackage{amsmath}
\usepackage{amssymb}
\usepackage{amsxtra}
\usepackage{physics}
\usepackage{mathrsfs}
\usepackage{booktabs}
\usepackage{gensymb}
\usepackage{graphicx}
\graphicspath{ {./images/} }
\usepackage{tikz}
\usepackage{hyperref}
\usepackage{lipsum}
\usepackage{comment}

\newcommand{\ii}{{i}}
\usepackage{mathtools}

\renewcommand{\Im}{{\rm Im}}

\newcommand{\red}[1]{{\color{black}{#1}}}

\begin{document}

\title{Time-diffracting 2D wave vortices}

\author{Boris A. Khanikati}
\affiliation{Department of Mathematics, Stony Brook University, Stony Brook, NY 11794, USA}

\author{Konstantin Y. Bliokh}
\affiliation{Donostia International Physics Center (DIPC), Donostia-San Sebasti\'{a}n 20018, Spain}
\affiliation{IKERBASQUE, Basque Foundation for Science, Bilbao 48009, Spain}
\affiliation{Centre of Excellence ENSEMBLE3 Sp.~z o.o., 01-919 Warsaw, Poland}

\begin{abstract}
Wave vortices constitute a large family of wave entities, closely related to phase singularities and orbital angular momentum (OAM). So far, two main classes of localized wave vortices have been explored: (i) transversely-localized monochromatic vortex beams that carry well-defined longitudinal OAM and propagate/diffract \red{along the longitudinal $z$-axis in space}, and (ii) 2D-localized spatiotemporal vortex pulses that carry the more elusive transverse (or tilted) OAM and propagate/diffract \red{along both the $z$-axis and time}. Here we introduce another class of wave vortices which are localized in a 2D $(x,y)$ plane, do not propagate in space (apart from uniform radial deformations), and instead propagate/diffract \red{solely along time}. These vortices possess well-defined transverse OAM and can naturally appear in 2D wave systems, such as surface polaritons or water waves. We provide a general integral expression for time-diffracting 2D wave vortices, their underlying ray model, and examples of approximate and exact wave solutions. We also analyze the temporal Gouy phase closely related to the rotational evolution in such vortices.
\red{Finally, we show that time-diffracting 2D vortices can provide strong spatiotemporal concentration of energy and OAM at sub-wavelength and oscillation-period scales.}
\end{abstract}

\maketitle 

\section{Introduction}

Wave vortices are remarkable physical objects that appear in wave fields of various natures, including optical \cite{Allen_book, Bekshaev_book, Torres_book}, acoustic \cite{Hefner1999JASA, Lekner2006JASA, Volke-Sepulveda2008PRL}, quantum matter waves \cite{Bliokh2017PR, Clark2015Nature, Luski2021Science}, and water waves \cite{Smirnova2024PRL, Wang2024}. 
The characteristic feature of a wave vortex is the presence of a phase singularity: a nodal point where the field intensity vanishes, while the phase acquires a $2\pi\ell$ increment upon encircling it, with $\ell$ being the integer vortex charge \cite{Nye1974}. Moreover, circularly-symmetric vortices carry a well-defined orbital angular momentum (OAM) associated with this winding phase \cite{Ceperley1992AJP, Allen1992PRA, Allen_book, Bekshaev_book, Torres_book, Hefner1999JASA, Lekner2006JASA, Volke-Sepulveda2008PRL, Bliokh2017PR, Clark2015Nature, Luski2021Science}. 

\begin{figure}[t]
\centering
\includegraphics[width=\linewidth]{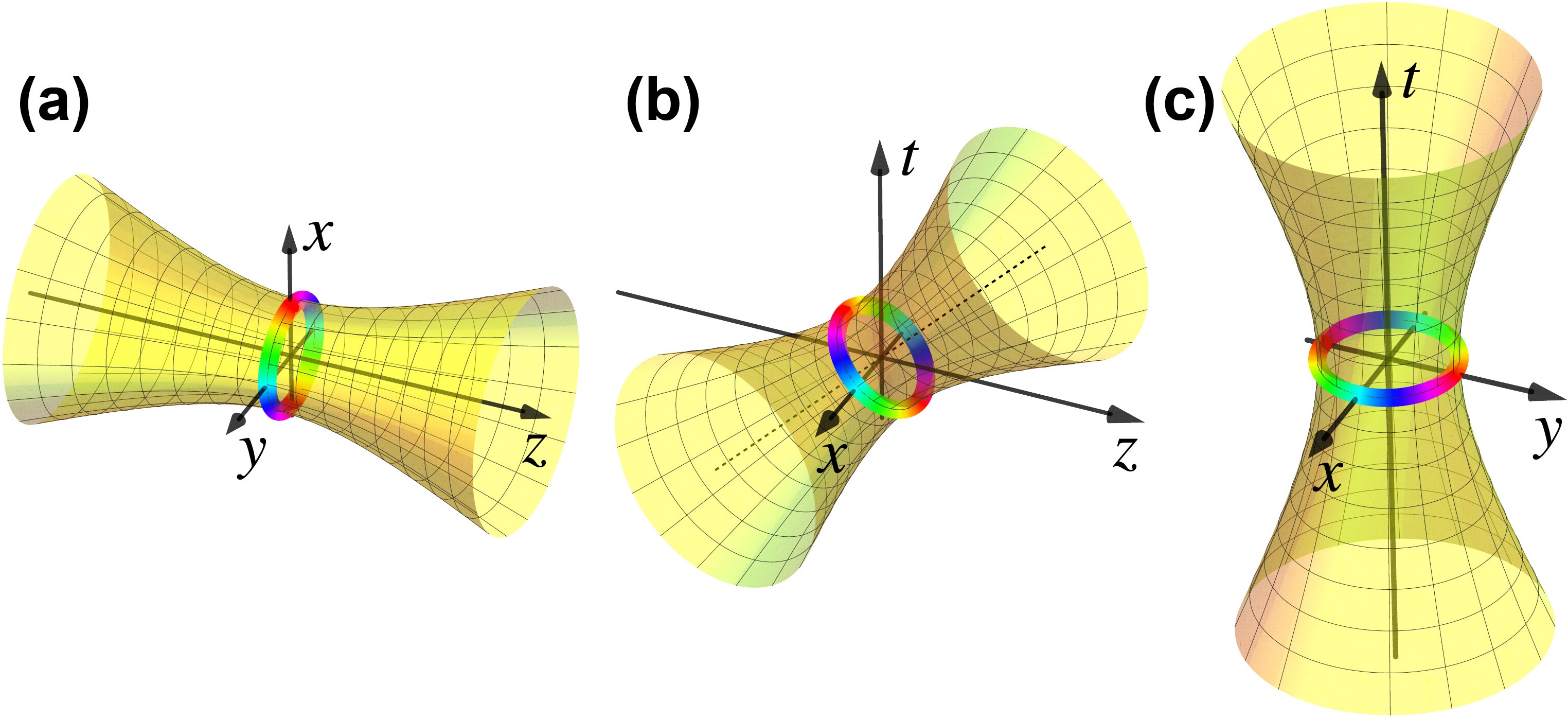}
\caption{\red{Schematics of (a) monochromatic spatial vortex beams propagating and diffracting along the $z$-axis; (b) spatiotemporal vortices propagating and diffracting along the $z$-axis and time $t$; and (c) 2D vortices propagating and diffracting in time $t$, which are the focus of this work.}}
\label{fig:intro}
\end{figure}

The widely explored {\it vortex beams} are solutions of wave equations that are localized in the transverse $(x,y)$ plane, while propagating and diffracting along the longitudinal $z$-axis \cite{Allen_book, Bekshaev_book, Torres_book, Hefner1999JASA, Lekner2006JASA, Volke-Sepulveda2008PRL, Bliokh2017PR, Clark2015Nature, Luski2021Science}, \red{see Fig.~\ref{fig:intro}(a)}. These beams are monochromatic, exhibiting harmonic time dependence and a stationary intensity distribution. For circularly-symmetric beams, the phase singularity resides on the beam axis, and the corresponding intrinsic OAM is directed along the $z$-axis.

Recently, a new class of {\it spatiotemporal wave vortices} has attracted considerable attention \cite{Sukhorukov2005, Bliokh2012, Jhajj2016, Chong2020, Zhang2023NC, Che2024PRL}. In the simplest configuration, such vortices are confined within the $(x,z)$ plane, while propagating and diffracting along the $z$-axis. These vortices are essentially polychromatic and time-varying: they propagate and diffract both along the $z$-axis and time $t$, and their phase singularities manifest in both the $(x,z)$ and $(x,t)$ planes, \red{see Fig.~\ref{fig:intro}(b)}. The corresponding vortex-induced OAM is directed along the orthogonal $y$-axis; however, its theoretical description has been a subject of debate \cite{Hancock2021, Bliokh2023PRA, Porras2024JO}, because of the intrinsically asymmetric (anisotropic) structure of these vortices. Furthermore, this transverse OAM is somewhat elusive, as its value depends sensitively on the choice of the coordinate origin. 

Both monochromatic (spatial) and spatiotemporal vortices can occur in 2D wave systems. For example, monochromatic vortices have been observed in surface plasmon-polaritons \cite{Ohno2006OE, Gorodetski2008, Prinz2023} and water-surface waves \cite{Smirnova2024PRL, Wang2024}. However, monochromatic solutions of the 2D wave equation cannot be truly localized in an infinite homogeneous plane, as they are not square-integrable and therefore possess infinite energy. Hence, such vortices must be spatially bounded and can be generated only in finite-size cavities (or outside holes in the plane \cite{Domina2025}). In turn, spatiotemporal 2D vortices can be localized, but they propagate and diffract in both space and time, such that their vortex structure typically degrades away from the focal region. 

In this work, we describe another class of 2D wave vortices that are spatially localized (i.e., possess finite energy), do not propagate in space (i.e., have a stationary phase singularity and well-defined intrinsic OAM), and propagate/diffract only along time, \red{as shown in Fig.~\ref{fig:intro}(c)}. These vortices can be regarded as spatiotemporal analogues of monochromatic vortex beams, where the propagation $z$-axis is replaced by time $t$, as well as a limiting case of spatiotemporal vortices, for which the propagation velocity vanishes. We present the general integral expression as well as examples of exact-analytical and approximate-model solutions for such time-diffracting 2D wave vortices. In addition, we examine the Gouy phase associated with their temporal diffraction and rotational dynamics. 
\red{Remarkably, time-diffracting 2D vortices can provide strong spatiotemporal concentration of energy and OAM at sub-wavelength scales, thereby offering a promising tool for linear and nonlinear optical interactions.}

\section{General approach}

We consider scalar 2D waves described by a complex wavefunction $\Psi({\bf r},t)$, ${\bf r}=(x,y)$, which can be represented as a superposition of plane waves $e^{i {\bf k}\cdot{\bf r}-i\omega t }$ with wavevector ${\bf k}$ and frequency $\omega$ determined by an isotropic dispersion relation $\omega(k)$.

\begin{figure}[t]
\centering
\includegraphics[width=\linewidth]{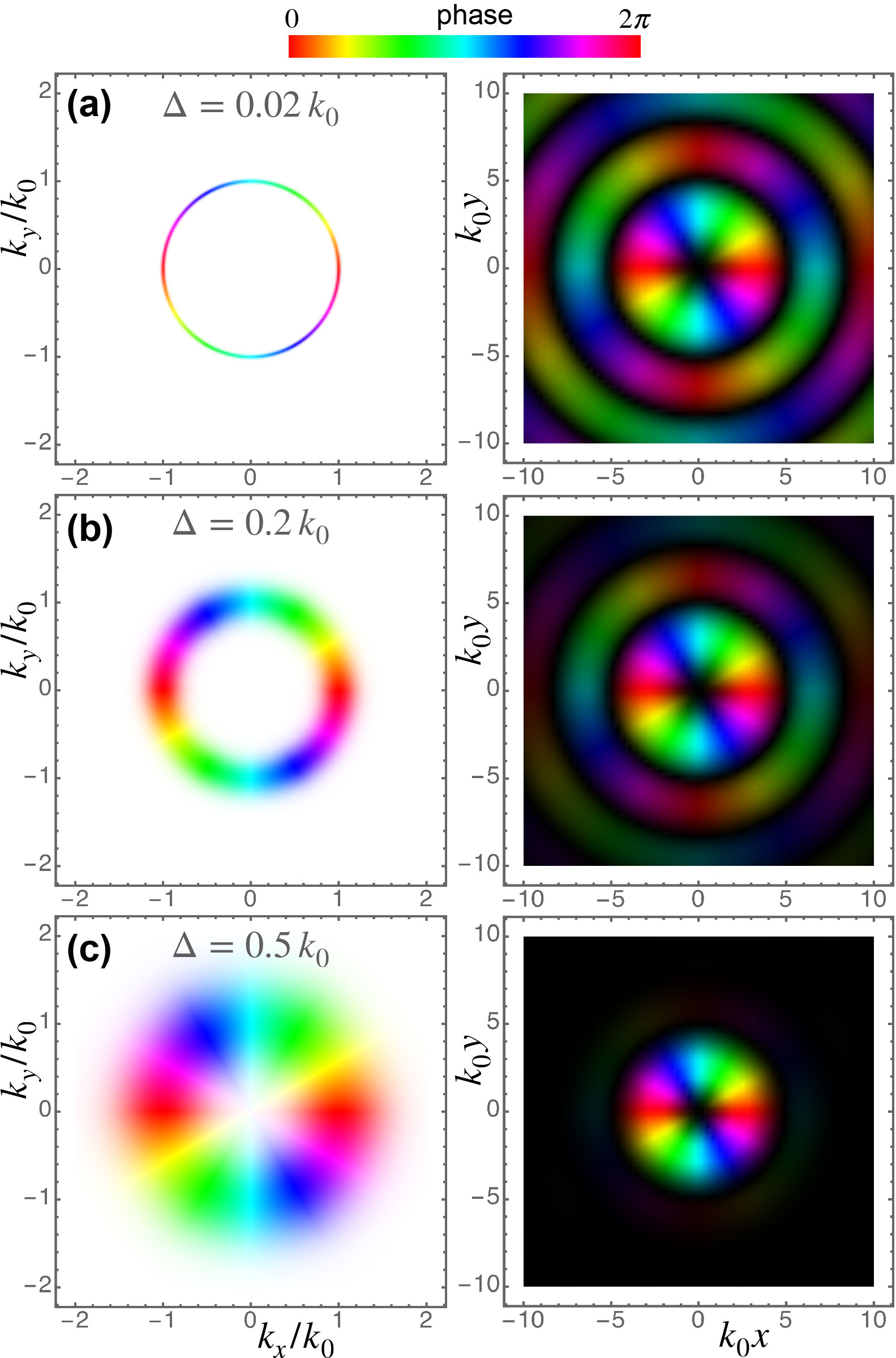}
\caption{Examples of 2D wave vortices \eqref{eq:vortex} for $\ell=2$ and wavenumber distributions $f(k) = \exp\left[-(k-k_0)^2/\Delta^2 \right]$ with different values of $\Delta$. Left column: plane-wave spectra in the ${\bf k}$-plane, where saturation and colors represent the amplitude and phase of the plane-wave components. Right column: real-space distributions of $\Psi({\bf r},t=0)$, where brightness and colors represent the amplitude and phase of the wavefunction.
The limit $\Delta\to 0$, shown in (a), corresponds to the monochromatic Bessel vortex \eqref{eq:Bessel} with multiple intensity rings. For $\Delta \sim k_0$, shown in (c), the vortex becomes practically localized within a single intensity ring.}
\label{fig:vortex}
\end{figure}

Monochromatic 2D vortices can be constructed as superpositions of plane waves with the same frequency $\omega_0= \omega(k_0)$, wavevectors uniformly distributed over the circle ${\bf k}=k_0(\cos\phi,\sin\phi)$ in ${\bf k}$-space, with azimuthal angle $\phi \in [0,2\pi)$, and additional phases $e^{i\ell\phi}$, $\ell \in \mathbb{Z}$, Fig.~\ref{fig:vortex}(a). This results in 
\begin{align}
\label{eq:Bessel}
\Psi_B \propto \int_0^{2\pi} e^{i\vb{k}\cdot\vb{r}-i\omega_0 t+i\ell\phi}\,d\phi 
\propto J_{|\ell|}(k_0r)e^{i \ell\varphi -i\omega_0 t},
\end{align}
where $J_{|\ell|}$ is the Bessel function of the first kind, and $\varphi$ is the azimuthal angle in real space such that ${\bf r}= r(\cos\varphi,\sin\varphi)$ and ${\bf k}\cdot {\bf r} = kr \cos(\varphi-\phi)$. Constant amplitude factors are omitted for simplicity.

Equation~\eqref{eq:Bessel} describes the well-known Bessel-type vortices with an infinite number of radial intensity maxima (rings) whose amplitudes decay as $\propto 1/\sqrt{r}$ at $r \to \infty$, Fig.~\ref{fig:vortex}(a). Such vortices can be realized in limited circular cavities containing a finite number of Bessel maxima \cite{Ohno2006OE, Gorodetski2008, Wang2024}, but cannot be extended to the entire plane. Indeed, the integral $\iint_{\mathbb{R}^2} |\Psi_B|^2\, r\, dr\,d\varphi \propto \int_0^\infty |J_{|\ell|}(kr)|^2\, r\,dr$ diverges, 
indicating that these solutions possess infinite total energy. 

This behavior can be understood from the uncertainty principle. The plane-wave spectrum of a Bessel vortex lies on a 1D circular contour in ${\bf k}$-space, characterized by a delta-function radial distribution $\delta(k-k_0)$. Therefore, the field becomes radially delocalized in the conjugate ${\bf r}$-space.

\begin{figure*}[t]
\centering
\includegraphics[width=\linewidth]{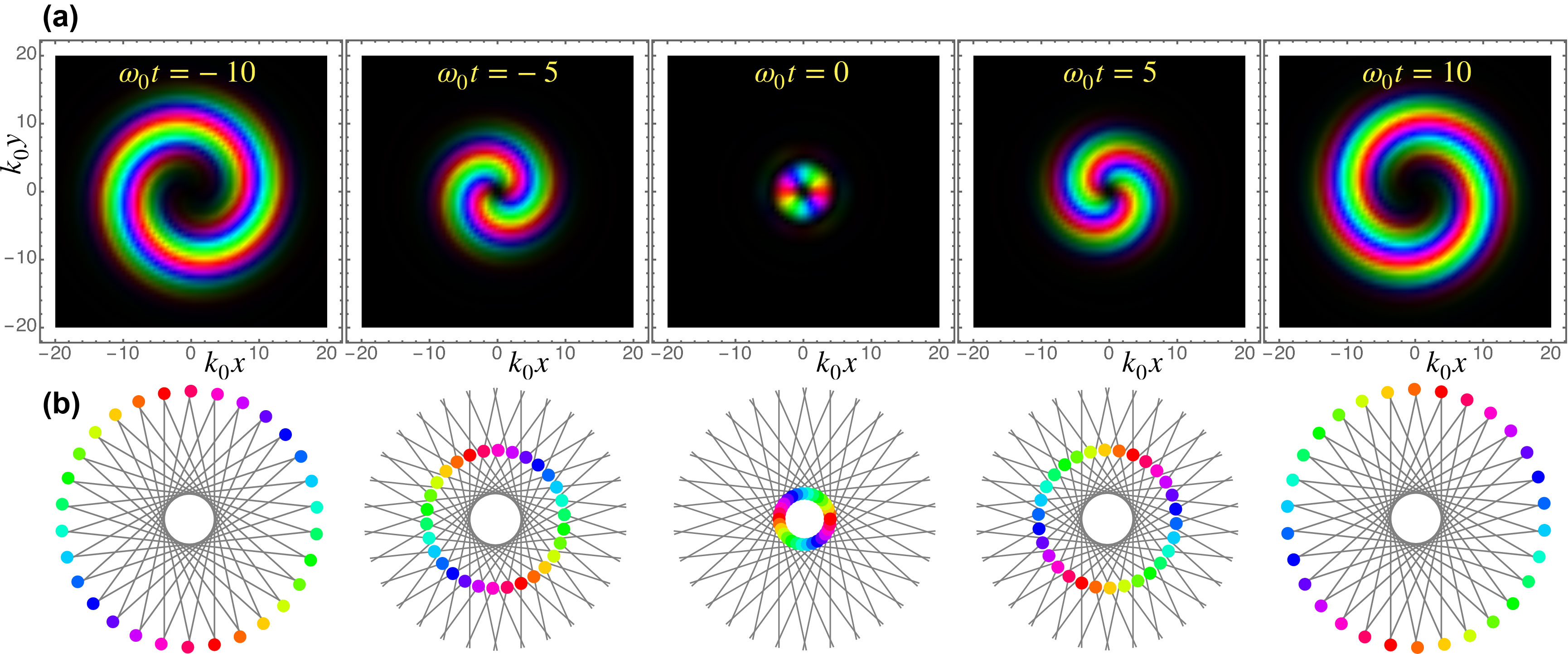}
\caption{(a) Temporal evolution of the vortex from Fig.~\ref{fig:vortex}(c) for waves with linear dispersion $\omega = kc$. \red{In each panel, the intensity is normalized by its maximum value (for the evolution of intensity with time see Fig.~\ref{fig:radial}).} (b) Geometrical-optics rays \eqref{eq:rays} and the corresponding moving point particles underlying the temporal diffraction shown in (a). 
}
\label{fig:diffraction}
\end{figure*}

To achieve localization of the vortex in the 2D plane, one must consider a plane-wave spectrum with a finite-width distribution in $k$, Fig.~\ref{fig:vortex}(b). 
This necessarily produces a polychromatic spectrum with frequencies $\omega(k)$. Introducing such a distribution $f(k)$ on top of the azimuthal plane-wave distribution of the Bessel vortices \eqref{eq:Bessel}, we obtain:
\begin{equation}
\label{eq:vortex}
\Psi \propto e^{i\ell\varphi} \int_0^\infty f(k) J_{|\ell|}(kr) e^{-i\omega(k) t}\,k \, dk \, .
\end{equation}
Equation~\eqref{eq:vortex} represents the general form of a circularly-symmetric localized 2D wave vortex $\Psi({\bf r},t)$, see Figs.~\ref{fig:vortex}(b,c). \red{From Parseval's theorem (i.e., equivalence of the wavefunction norm in the $r$- and $k$-representations),} its total energy is proportional to 
\begin{equation}
\label{eq:energy}
\int_0^\infty |\Psi|^2\, r\, dr \propto \int_0^\infty |f(k)|^2\,k\, dk\,,
\end{equation}
and is finite provided that the radial spectrum $f(k)$ is square-integrable.

Owing to the global azimuthal phase factor $e^{i\ell \varphi}$, this vortex exhibits a stationary phase singularity of order $\ell$ at ${\bf r}={\bf 0}$, and carries a well-defined OAM with respect to the orthogonal $z$-axis: 
\begin{equation}
\label{eq:OAM}
\hat{L}_z \Psi = \ell\, \Psi\,,
\end{equation}
where $\hat{L}_z = -i \partial/\partial\varphi$ is the quantum-mechanical OAM operator (in units $\hbar =1$) \cite{Allen_book}. Equation~\eqref{eq:OAM} implies that the vortex \eqref{eq:vortex} has an expectation (integral) value of OAM $\langle L_z \rangle = \ell$, normalized ``per particle'' \cite{Allen_book, Bekshaev_book, Torres_book, Hefner1999JASA, Lekner2006JASA, Volke-Sepulveda2008PRL, Bliokh2017PR}. Notably, this OAM is purely {\it intrinsic}, i.e., independent of the choice of coordinate origin \cite{Berry1998}. Indeed, under a shift of the coordinate origin, ${\bf r} \to {\bf r} + {\bf a}$, the expectation value of the OAM transforms as $\langle {\bf L} \rangle \to \langle {\bf L} \rangle + {\bf a} \times \langle {\bf p}\rangle$, where $\langle {\bf p}\rangle$ is the expectation value of the wave momentum \cite{Bliokh2015PR}. This relation shows that the longitudinal OAM component, parallel to $\langle {\bf p}\rangle$, remains invariant, as in the case of monochromatic vortex beams. For the vortices \eqref{eq:vortex}, the mean momentum vanishes due to the circularly-symmetric distribution of interfering plane waves, $\langle {\bf p}\rangle = {\bf 0}$, making the integral OAM origin-independent.   

Figure~\ref{fig:vortex} shows examples of localized 2D vortices \eqref{eq:vortex} with $\ell=2$ and distribution $f(k) = \exp\left[-(k-k_0)^2/\Delta^2 \right]$ (strictly speaking, it should satisfy $f(0)=0$, but it has an exponentially small contribution from $f(0)$). 
For $\Delta \ll k_0$, the vortex resembles the Bessel vortex \eqref{eq:Bessel} with multiple intensity rings. As $\Delta$ increases, the amplitudes of the outer rings diminish, and for $\Delta \sim k_0$, the energy becomes largely confined to the innermost ring, similar to the Laguerre-Gaussian ``donut'' beams widely used across various domains of wave physics \cite{Allen_book, Bekshaev_book, Torres_book, Hefner1999JASA, Lekner2006JASA, Volke-Sepulveda2008PRL, Bliokh2017PR, Clark2015Nature, Luski2021Science}.

The price of spatial localization is the inherently polychromatic, non-stationary character of the vortex \eqref{eq:vortex}. Figure~\ref{fig:diffraction} shows an example of the temporal evolution of such vortices. 
One can see that this evolution parallels the $z$-propagation and diffraction of transversely-localized vortex beams \cite{Allen_book, Bekshaev_book, Torres_book, Hefner1999JASA, Lekner2006JASA, Volke-Sepulveda2008PRL, Bliokh2017PR, Clark2015Nature, Luski2021Science}, see Fig.~\ref{fig:intro}. For real-valued distributions $f(k)$, the instant $t=0$ corresponds to the `temporal focal plane', characterized by the minimal vortex radius, \red{maximum concentration of energy (see Fig.~\ref{fig:radial} below),} and a purely azimuthal phase gradient. At times $|\omega_0 t| \gg 1$, the field reaches the far-field regime, where the vortex radius grows near-linearly with $|t|$ and the phase gradient becomes predominantly radial (while maintaining the $2\pi\ell$ phase increment around the vortex core). Thus, the time-diffracting vortices described by Eq.~\eqref{eq:vortex} represent spatio-temporal counterparts of monochromatic vortex beams diffracting upon spatial propagation. 

\section{Approximate and exact solutions}

\subsection{Geometrical optics ray picture}

\begin{figure*}[t]
\centering
\includegraphics[width=\linewidth]{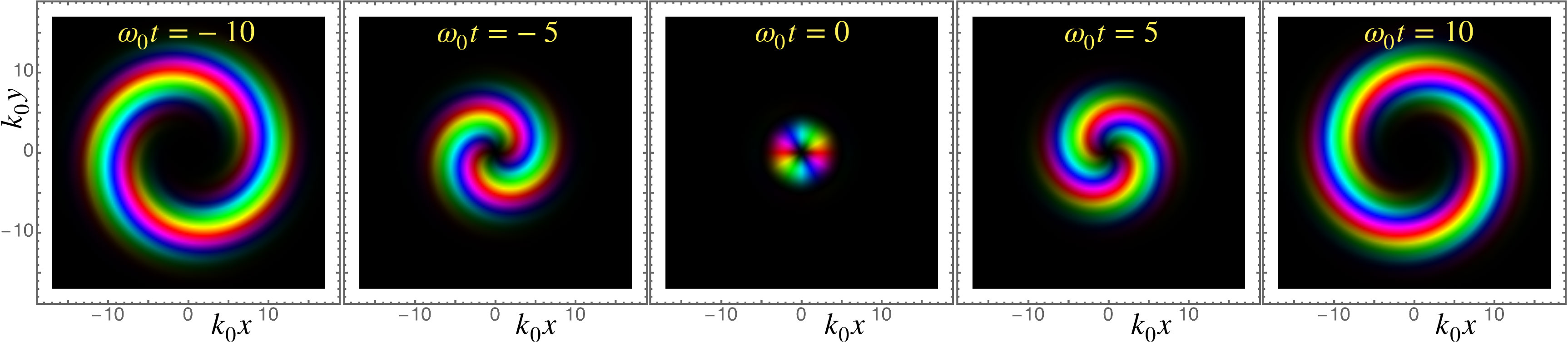}
\caption{Temporal evolution of the vortex described by a simple analytical model \eqref{eq:model} with $\ell=2$ and $k_0L =3$.
}
\label{fig:model}
\end{figure*}

The temporal diffraction of vortices \eqref{eq:vortex}, confined to a single intensity ring, can be understood using a simple geometrical-optics ray model. Consider a family of rectilinear rays described by equation
\begin{equation}
\label{eq:rays}
{\bf r}(\phi,t) = {\bf r}_0 + {\bf u}t \equiv
r_0\left(\cos\phi,\sin\phi \right)+\sigma u\left(-\sin\phi,\cos\phi \right)t\,,
\end{equation}
where $\phi \in [0,2\pi)$, $\sigma={\rm sgn}(\ell)$, $r_0 = |\ell|/k_0$, and $u=\partial\omega /\partial k$ is the group velocity of the waves.
These rays are tangent to a circular caustic of radius $r_0$, corresponding to the vortex radius at the focal time $t=0$ ($k_0 r_0 = |\ell|$ serves as the appropriate quantization condition). Now, consider point particles moving along these rays according to Eq.~\eqref{eq:rays}, each marked by the vortex phases $\ell\phi$. As illustrated in Fig.~\ref{fig:diffraction}(b), this classical rectilinear motion mimics the temporal evolution of the wave vortex with very good accuracy. 

\subsection{Model wave solution}

One can also construct a model wave solution describing the temporal diffraction of the confined vortices \eqref{eq:vortex}.
Instead of interfering plane waves propagating at different azimuthal angles $\phi$ and forming the Bessel vortex \eqref{eq:Bessel}, we consider a similar interference of Gaussian {\it wavepackets}. Assuming linear dispersion, $\omega = kc$, the wavefield of an individual wavepacket can be written as   
\begin{equation}
\psi \propto e^{-\frac{r^2\sin^2(\varphi-\phi)}{w^2}} 
e^{-\frac{[r\cos (\varphi-\phi) - ct]^2}{L^2}} e^{\ii k_0[r\cos(\varphi-\phi)-c t]+i\ell\phi}.
\label{eq:wp}
\end{equation}
Here, $w$ and $L$ are the width and length of the wavepackets, $k_0$ is the central wavenumber, and diffraction of individual packets is neglected. 

The resulting wavefield is $\Psi_{\rm model} = \int_0^{2\pi} \psi\, d\phi$. This integral cannot be evaluated analytically for the general wavepackets \eqref{eq:wp}, but can be computed for the special case $w=L$. This yields
\begin{equation} 
\label{eq:model}
\Psi_{\rm model} \propto e^{i \ell \varphi-\ii \omega_0 t} e^{-\frac{r^2+c^2t^2}{L^2}} J_{|\ell|}\! \left(k_0r - i\frac{2ctr}{L^2} \right),
\end{equation}
where $\omega_0= k_0c$.
The parameter $L$ controls the thickness and number of vortex rings. For $k_0 L\lesssim 1$, the wavefunction \eqref{eq:model} is essentially confined to a single intensity ring. For $k_0L \gg 1$, additional rings emerge, and in the limit $L\to \infty$ the wavefunction reproduces the monochromatic Bessel vortex \eqref{eq:Bessel}. This behavior is similar to the transition from $\Delta\sim 1$ to $\Delta \ll 1$ in Fig.~\ref{fig:vortex}

Figure~\ref{fig:model} demonstrates that the model wavefunction \eqref{eq:model} accurately captures the temporal evolution of the vortex \eqref{eq:vortex}, shown in Fig.~\ref{fig:diffraction}, up to an overall time-dependent amplitude factor. 

\subsection{An exact solution}

\begin{figure*}[t]
\centering
\includegraphics[width=\linewidth]{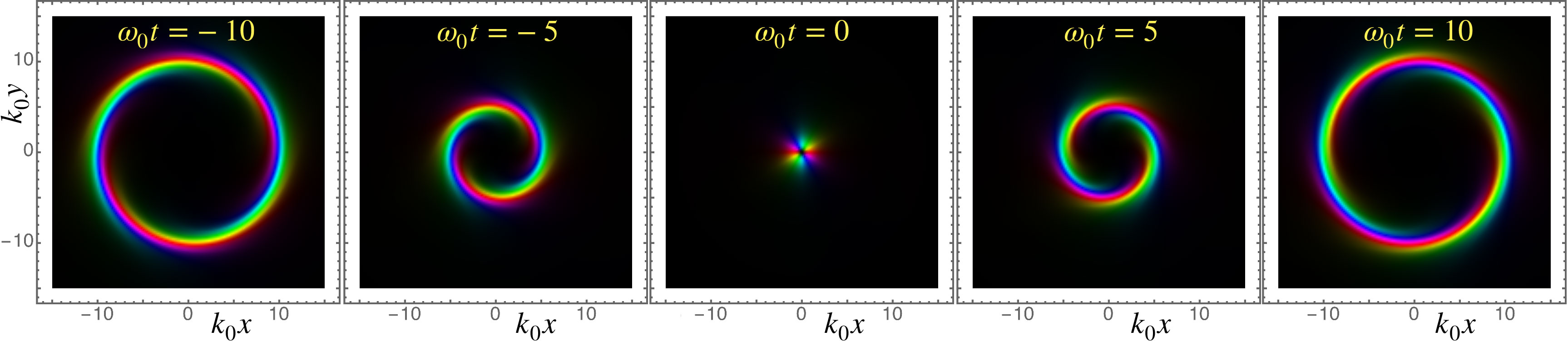}
\caption{Temporal evolution of the vortex described by the exact solution \eqref{eq:exact} with $\ell=2$.
}
\label{fig:exact}
\end{figure*}

Finally, for linear dispersion, $\omega=kc$, the general integral form \eqref{eq:vortex} can be evaluated analytically for the distribution
\begin{equation}
\label{eq:distribution}
f(k) = k e^{-k/k_0},
\end{equation}
which attains a maximum at $k = k_0$. The corresponding vortex wavefunction is given by:
\begin{align}
\label{eq:exact}
&\Psi \propto e^{\ii \ell \varphi} \frac{(k_0 r)^{|\ell|}}{(1 + \ii \omega_0 t)^{|\ell|+3}} \nonumber\\ 
&\times {}_2F_1\!\left(\frac{{|\ell|+3}}{2}, \frac{{|\ell|+4}}{2}; |\ell|+1; -\frac{k_0^2r^2}{(1+\ii \omega_0 t)^2} \right),
\end{align}
where $\omega_0 = k_0 c$ and ${}_2F_1$ is the hypergeometric function. Since the distribution \eqref{eq:distribution} is relatively broad, the vortex \eqref{eq:exact} is confined to a narrow intensity ring, as shown in Fig.~\ref{fig:exact}. 

\section{Temporal Gouy phase}

Diffraction of monochromatic vortex beams reveals the {\it Gouy phase} \cite{Feng2001}, which is closely related to the rotational evolution of such beams \cite{Basistiy1993OC, Arlt2003JMO, Hamazaki2006OE, Baumann2009OE, Cui2012JO, Guzzinati2013PRL, Schattschneider2014NC, Ghosh2024JOSA}. Similarly, the temporal diffraction of localized 2D vortices induces a temporal Gouy phase. This phase is responsible for the intrinsic azimuthal rotation by an angle $\pi\, {\rm sgn}(\ell)$ between the far-field limits $t\to -\infty$ and $t\to \infty$. Because of the $e^{i\ell\varphi}$ factor, this rotation corresponds to an additional Gouy phase
\begin{equation}
\label{eq:Gouy}
\Phi_G = \pi (|\ell|+1)\,.
\end{equation}

The simplest explanation of this effect follows from the ray picture in Eq.~\eqref{eq:rays} and Fig.~\ref{fig:diffraction}(b). It is easy to see that each particle moving along an infinite ray changes its azimuthal coordinate from $\varphi_i$ at $t\to -\infty$ to $\varphi_i + \pi\, {\rm sgn}(\ell)$. The same rotation can be observed in the temporal evolution of wavefunctions in Figs.~\ref{fig:diffraction}(b), \ref{fig:model}, and \ref{fig:exact} by tracking the azimuthal position of a point of a chosen color (phase) at the radial amplitude maximum. For instance, the red point with azimuthal position $\varphi = -\pi/2$ at $t\to -\infty$ moves to $\varphi=0$ at $t=0$ and then to $\varphi = \pi/2$ at $t\to \infty$.

The azimuthal rotation corresponding to the Gouy phase can also be derived from the general form of the vortex \eqref{eq:vortex} using the method developed in \cite{Schattschneider2014NC, Ghosh2024JOSA}. We assume a ``relativistic'' dispersion $\omega = \sqrt{c^2k^2+\mu^2}$, for which $\partial \omega /\partial k = c^2k/\omega$. The corresponding wave equation is the Klein-Gordon equation (which reduces to the usual wave equation when $\mu=0$). For this equation, the local probability current is given by ${\bf j}={\rm Im}\!\left(\Psi^*{\boldsymbol{\nabla}}\Psi\right)$, whereas the probability density is $\rho=-\Im\!\left(\Psi^*\partial\Psi\right/\partial t)$ \cite{BLP}. 

Then, the local velocity inside the vortex can be defined as ${\bf v} = c^2 {\bf j}/\rho$, and the local angular velocity is determined by its azimuthal component: $\Omega = v_\varphi/r$. Substituting the wavefunction \eqref{eq:vortex}, we obtain
\begin{equation}
\label{eq:Omega}
\Omega = \frac{c^2\ell}{r^2}\frac{|\Psi|^2}{\rho}\,.
\end{equation}
Next, we define the mean angular velocity, averaged over the radial distribution of the vortex: 
\begin{equation}
\label{eq:Omega-mean}
\langle\Omega\rangle =\frac{\int_0^\infty \rho\, \Omega\,r\,dr}{\int_0^\infty \rho\,r\,dr} = - c^2\ell\,\frac{\int_0^\infty r^{-1}|\Psi|^2\,dr}{\int_0^\infty \Im\!\left(\Psi^*\partial\Psi\right/\partial t)\,r\,dr}\,.
\end{equation}
Finally, the integral rotation of the vortex over the temporal evolution $t\in (-\infty, \infty)$ is determined by
\begin{equation}
\label{eq:rotation}
\Delta\varphi=\int_{-\infty}^{\infty}\langle\Omega\rangle\,dt\,.
\end{equation}

Substituting Eq.~\eqref{eq:vortex} into Eq.~\eqref{eq:Omega-mean}, we find that its denominator is equal to $\int_0^\infty \rho\,r\,dr = \int_0^\infty \omega(k) |f(k)|^2\,k\,dk$, which is time-independent. Next, substituting Eq.~\eqref{eq:vortex} into Eq.~\eqref{eq:Omega-mean} and performing the time integration \eqref{eq:rotation}, similarly to \cite{Ghosh2024JOSA}, yields 
\begin{equation}
\label{eq:rotation-result}
\Delta\varphi=\pi \, {\rm sgn}(\ell)\,.
\end{equation}

This is the exact desired result. It means that any azimuthal asymmetry or defect in the vortex \eqref{eq:vortex} undergoes rotation by the angle \eqref{eq:rotation-result}. This rotation can be observed experimentally, analogous to spatial Gouy-phase measurements \cite{Basistiy1993OC, Arlt2003JMO, Hamazaki2006OE, Baumann2009OE, Cui2012JO, Guzzinati2013PRL, Schattschneider2014NC, Ghosh2024JOSA}.

\section{Conclusions}

To summarize, we have introduced a novel class of 2D wave vortices that are localized in the $(x,y)$ plane but diffract in time. Unlike recently considered spatiotemporal vortices \cite{Sukhorukov2005, Bliokh2012, Jhajj2016, Chong2020, Zhang2023NC, Che2024PRL, Hancock2021, Bliokh2023PRA, Porras2024JO}, the center of these vortices is stationary, and the OAM with respect to the transverse $z$-axis is well defined. The temporal evolution of such vortices strongly resembles spatial propagation of monochromatic vortex beams \cite{Allen_book, Bekshaev_book, Torres_book, Hefner1999JASA, Lekner2006JASA, Volke-Sepulveda2008PRL, Bliokh2017PR, Clark2015Nature, Luski2021Science}, with the longitudinal $z$-axis replaced by time. 

These 2D localized vortices can naturally appear in surface-wave or planar-waveguide systems, such as surface polaritons, \red{surface acoustic waves,} or water-surface waves. \red{We have verified numerically that variations in dispersion relation $\omega(k)$ of these surface waves do not affect the main conclusions of this work.} \red{Furthermore, although such surface waves are vector waves, our scalar model can describe one component of the wave (e.g., the vertical electric field or the water-surface elevation, which are typically accessed experimentally); the remaining field components can then be reconstructed from the corresponding wave equations \cite{Tsesses2018S, Wang2024}.} 

We have presented a geometric-optics ray model underlying temporal diffraction of 2D vortices, a simple model solution based on Gaussian wavepacket interference, and an example of an exact analytical solution. The ray and model approaches provide a practical scheme for the generation of such vortices in 2D wave systems. Namely, one can adapt existing methods for generating monochromatic Bessel vortices in 2D circular cavities \red{with sources distributed along the circle and having azimuthal phase increment $\ell\varphi$} \cite{Ohno2006OE, Gorodetski2008, Wang2022JPC, Prinz2023, Wang2024} by replacing monochromatic sources with finite-length Gaussian-like pulses. The relative phase between pulses propagating in different directions sets the vortex strength $\ell$. In the far-field, the pulses form converging and diverging wave rings, while at the focal zone they interfere to produce a vortex with purely azimuthal phase gradient, as shown in Figs.~\ref{fig:diffraction}--\ref{fig:exact}.

Finally, we have analyzed the temporal Gouy phase, which manifests as an azimuthal rotation of $\pi\, {\rm sgn}(\ell)$ between the far-field zones. This rotation can be directly observed in experiments by tracking azimuthal defects in 2D vortices. Overall, the vortices described here extend the family of wave vortices, complementing previously known monochromatic and spatiotemporal vortices.

\begin{figure}[t]
\centering
\includegraphics[width=\linewidth]{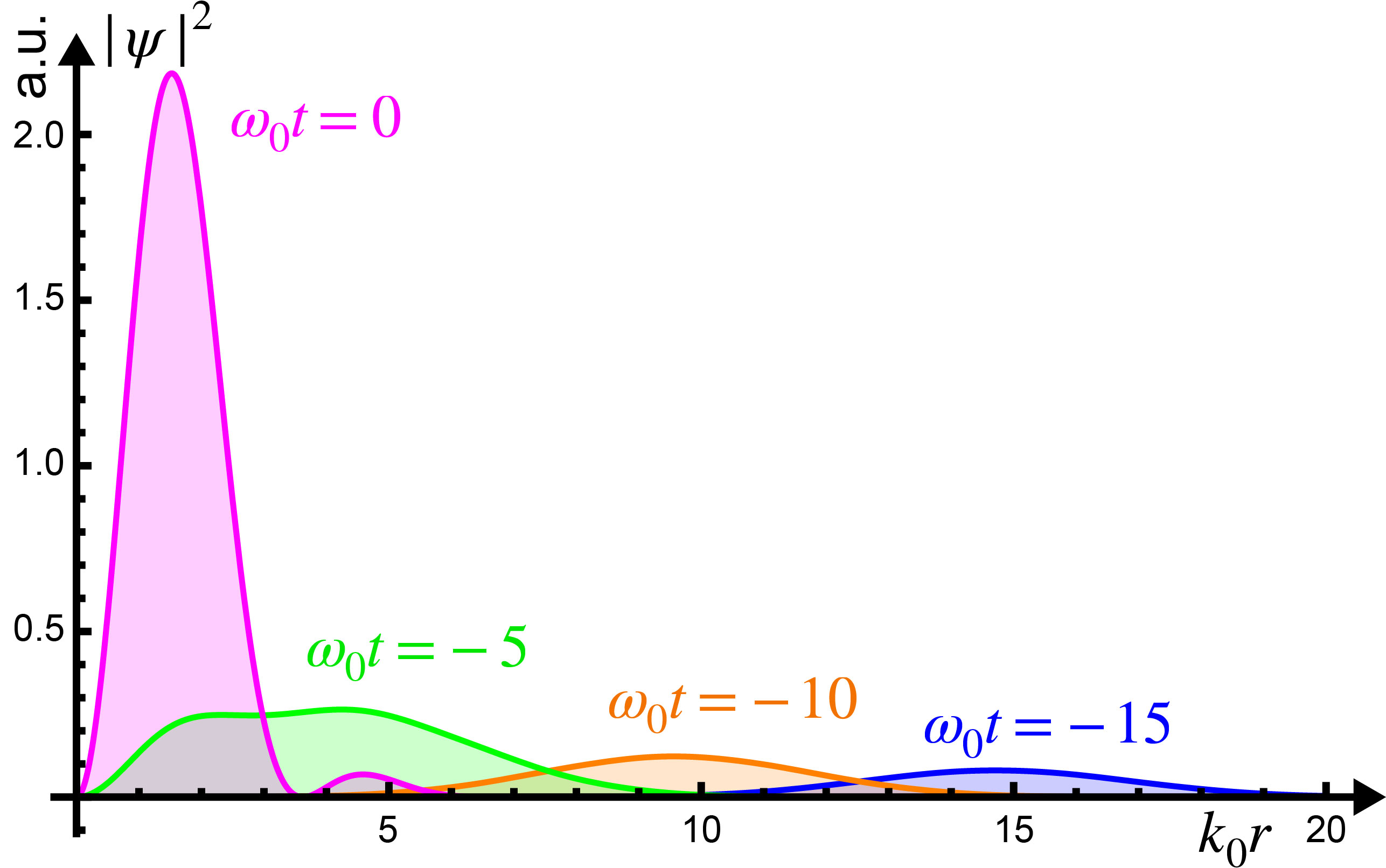}
\caption{\red{Radial intensity profiles of the vortex wavefunction, such as shown in Fig.~\ref{fig:diffraction} but with $\ell=1$, at different instants of time. The vortex energy is strongly concentrated at the sub-wavelength and oscillation-period scales in space and time.}}
\label{fig:radial}
\end{figure}

\red{Importantly, the vortices described here can be accompanied by strong concentration of energy in space and time. Figure~\ref{fig:radial} shows the radial intensity profiles of the vortex wavefunction, such as shown in Fig.~\ref{fig:diffraction} but with $\ell=1$, at different instants of time. The maximum intensity occurs at 
the `temporal focal plane' $t=0$, when the vortex ring reaches its smallest radius. The maximum-intensity radius and half-width of the peak are about a quarter of the central wavelength: $r_{\rm max}\simeq \Delta r\simeq \lambda_0/4 = \pi/(2k_0)$.
The corresponding temporal half-width of this intensity peak is about the oscillation period, $\Delta t \simeq 2\pi/\omega_0 = T_0$. Such strongly confined, high-intensity vortices are promising for applications in optical high-harmonic generation \cite{Gariepy2014PRL, Martin-Hernandez2025NP}, light-matter interactions \cite{Ayuso2019NP, Fang2022LSA}, vortex lasers \cite{Miao2016S, Huang2020S}, and femtosecond spatiotemporal optics \cite{Shen2023}.} 

\section*{Acknowledgements}

We acknowledge helpful discussions with Prof. Miguel A. Alonso and Prof. Alexander Khanikaev,  
as well as support from Marie Sk\l{}odowska-Curie COFUND Programme of the European Commission (project HORIZON-MSCA-2022-COFUND-101126600-SmartBRAIN3), 
ENSEMBLE3 Project carried out within the International Research Agendas Programme (IRAP) of the Foundation for Polish Science co-financed by the European Union under the European Regional Development Fund (MAB/2020/14) and Teaming Horizon 2020 programme of the European Commission (GA. No. 857543), 
and Minister of Science and Higher Education ``Support for the activities of Centers of Excellence established in Poland under the Horizon 2020 program'' (contract MEiN/2023/DIR/3797).




\bibliography{mybib}

\end{document}